\documentstyle[aas2pp4]{article}

\slugcomment{ to appear in Ap. J. Letters, xx.}

\lefthead{McArthur et al.}
\righthead{Parallax of RW Triangulum  }

\begin{document}

\title{Astrometry with {\it Hubble Space Telescope} Fine Guidance Sensor 3: \\ The Parallax of the Cataclysmic Variable \\RW
Triangulum} 

\author{B. E. McArthur and G. F. Benedict}
\affil{McDonald Observartory, University of Texas, Austin, Texas 78712}
\author{J. Lee, C. -L. Lu \altaffilmark{1}, W. F. van Altena, C. P. Deliyannis \altaffilmark{2}, and  T. Girard}
\affil{Department of Astronomy, Yale University, New Haven, Connecticut 05620 }
\author{L. W. Fredrick}
\affil{Department of Astronomy, University of Virginia, Charlottesville, Virginia 22903}
\author{E. Nelan}
\affil{Space Telescope Science Institute, Baltimore, MD 21218}
\author{R. L. Duncombe, P. D. Hemenway \altaffilmark{3}, W. H. Jefferys and  P. J. Shelus}
\affil{Department of Astronomy, University of Texas, Austin, Texas 78712}
\author{O. G. Franz and L. H. Wasserman}
\affil{Lowell Observatory, Flagstaff, Arizona 86001}
\altaffiltext{1} {Now at Purple Mountain Observatory, Nanjing,China}
\altaffiltext{2} {Now at Department of Astronomy, Indiana University, Bloomington, Indiana}
\altaffiltext{3} {Now at Department of Oceanography, University of Rhode Island}

% Notice that each of these authors has alternate affiliations, which
% are identified by the \altaffilmark after each name.  The actual alternate
% affiliation information is typeset in footnotes at the bottom of the
% first page, and the text itself is specified in \altaffiltext commands.
% There is a separate \altaffiltext for each alternate affiliation
% indicated above.

% The abstract environment prints out the receipt and acceptance dates
% if they are relevant for the journal style.  For the aasms style, they
% will print out as horizontal rules for the editorial staff to type
% on, so long as the author does not include \received and \accepted
% commands.  This should not be done, since \received and \accepted dates
% are not known to the author.

\begin{abstract}
RW Triangulum (RW Tri)  is a 13th magnitude Nova-like Cataclysmic  Variable star
with an orbital period of 0.2319 days (5.56 hours). 
Infrared observations of RW Tri indicate that its secondary is most likely
a late K-dwarf (\cite{Dhi98}).
Past analyses predicted a distance of 270 parsec, 
derived from a black-body fit to 
spectrum of the central part of the disk (\cite{Rutt92}).
Recently completed Hubble Space Telescope  Fine Guidance
Sensor interferometric  observations  allow us to determine the first
trigonometric parallax  to RW Tri. This determination puts the distance
of RW Tri at $341^{-31}_{+38}$, one of the most distant objects with a 
direct parallax measurement.

We  compare our result with
methods previously employed to estimate distances to CV's.

\end{abstract}

% The different journals have different requirements for keywords.  The
% keywords.apj file, found on aas.org in the pubs/aastex-misc directory, 
% contains a list of keywords used with the ApJ and Letters.  These are 
% usually assigned by the editor, but authors may include them in their 
% manuscripts if they wish. 

\keywords{astrometry, stars: distances,stars: novae, cataclysmic variables}
%\keywords{globular clusters,peanut clusters,bosons,bozos}

% That's it for the front matter.  On to the main body of the paper.
% We'll only put in tutorial remarks at the beginning of each section
% so you can see entire sections together.

% In the first two sections, you should notice the use of the LaTeX \cite
% command to identify citations.  The citations are tied to the
% reference list via symbolic KEYs.  We have chosen the first three
% characters of the first author's name plus the last two numeral of the
% year of publication.  The corresponding reference has a \bibitem
% command in the reference list below.
%
% Please see the AASTeX manual for a more complete discussion on how to make
% \cite-\bibitem work for you.   

\section{Introduction}

Cataclysmic Variables (CVs) provide a rich library from which astronomers can
study physical phenomena.  Magnetic and plasma interactions, winds, 
non-equilibrium thermo-nuclear reactions, radiative emissions and accretion
can all be found in the laboratory of these white dwarfs and their
donor companions. All CVs are short-period binary systems
that transfer matter via    Roche-lobe overflow from a
red dwarf companion  to the white dwarf.
The Nova-Like (NL) CVs have accretion disks which remain bright at all times
making  it  difficult    to estimate their distances,
since the secondary is very difficult to observe.
The mass transfer rate of NLs is high enough to suppress the disc
instability mechanism that causes Dwarf Nova type outbursts.
RW Tri  is an eclipsing  NL star.   A precise parallax 
would increase our understanding of this class of object.
Estimates of the rates of mass 
transfer are required for quantitative modeling of the
emergent spectrum and evolution of these systems.
Only if the distance is known can accurate estimates
of the rates of mass transfer be made.

Berriman (1987) reviewed the four types of measurements used
to determine distances to CVs: 
(1) parallaxes and proper motions, (2) interstellar 
reddening, (3) properties of the accretion disks and (4) the 
detection of the red dwarf companions. 
Detection of the red dwarf companion of RW Tri has been detected 
spectroscopically by 
skew mapping (\cite {Smit93})  and  K-band spectroscopy (\cite{Dhi98}).
A fifth model-indpependent method
using linear polarimetry has recently been proposed  (\cite {Barr96}).
The different methods
give widely varying distances, and some measurements should be 
regarded as lower limits (e.g. when using the K-band magnitude
of the secondary star in conjunction with Bailey's relation -  see Section 4), or
upper limits (e.g. from blackbody disc models fitted to spectra or photometry).
Only recently has there been  instrumentation to improve the
astrometric database.  HIPPARCOS was  scheduled to observe 5 CVs
 (V603 Aql, RR Pic, RW Sex, SS Cyg and AE Aqr), but only
AE Aqr yielded a meaningful     parallax of 9.8{$\pm$}2.84 mas (\cite {Frie97}).
To date a total of four sub-milliarcsecond precision 
astrometric parallaxes of CVs have been delivered 
by the Fine Guidance Sensors(FGS) on
HST, three  dwarf novae SS Cyg, U Gem and SS Aur (\cite {Har99}), and
the NL RW Tri, which is the subject of this paper.

\section{Observations and Reductions}

The observations of RW Tri (J2000: 02 25 36.20, +28 05 50.5)
 were  made with Fine Guidance 
Sensor 3 (FGS3) on the HST.
Astrometry with the HST Fine Guidance
Sensors has been previously described (\cite{Ben94,Ben93}),
as has the FGS instrument (\cite{Bra91}).
Ten   observations (one orbit each) of RW Tri at maximum
parallax factor were made between 1995 and 
1998 with FGS3 in POS (fringe tracking) mode. 
HST FGS parallax observing strategies and reduction and analysis
techniques  have been
described  by  Benedict et. al. (1999), Harrisonet. al.  (1999)
and van Altena et. al. (1997).

\begin{figure}
\epsscale{.6}
\plotone{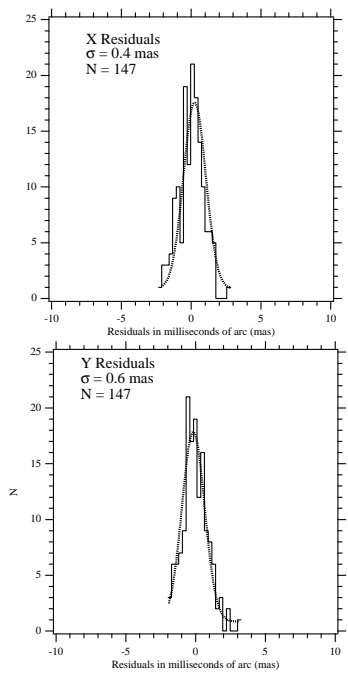}
\caption{Histograms of x and y residuals obtained from modeling  RW Tri 
and its reference frame.   Distributions are fit with 
Gaussians.} \label{fig-1}
\end{figure}

As seen in Table~\ref{tbl-2},
the standard errors  resulting from the solutions for relative
parallax and proper motion 
are sub-milliarcsecond.   Figure~\ref{fig-1}
shows histograms of the residuals of the target and reference frame
stars obtained from our astrometric modeling.
The proper motions and parallaxes determined with HST are relative to
the reference frame stars.  To determine the correction to absolute parallax,
spectrophotometric parallaxes were derived for the reference frame stars
r
from spectra obtained  at the
WIYN telescope\footnote{The WIYN Observatory is a joint 
facility of the University of
Wisconsin-Madison, Indiana University, Yale University, and the National
Optical Astronomy Observatories.}
~multiobject spectrograph (MOS/Hydra) and classified by spectral type and
luminosity class.
The final correction, 0.34{$\pm$}0.16 
milliseconds of arc (mas) is based upon $A_{v}$= 0.2.
This $A_{v}$ is an upper limit derived from Burstein and Heiles (1984),
using the galactic latitude of  RW Tri.

% Authors may indicate to the editorial staff where they would like 
% figures and tables to be placed in the manuscript.  This is done with
% either the \placefigure{KEY} or \placetable{KEY} commands.  These
% commands require \label{KEY} commands to be placed appropriately with
% corresponding table and figure captions.  When the manuscript is
% printed a short note is printed on the page where the figure or table
% is to go.  These commands are ignored in the aaspp4 and aas2pp4 styles.

%\placetable{tbl-3}
%\placefigure{fig1}

% In this section, we see the use of the \subsection command to set off
% an independent subsection.  We only have one here; usually there would
% be several.

% We show the use of several of the displayed math environments described
% in the User Guide, and you get a healthy dose of mathematical typesetting
% examples.  Also, observe the use of the LaTeX \label command after the
% \subsection to give a symbolic KEY to the subsection for cross-referencing
% in a \ref command.  LaTeX automatically numbers the sections, equations,
% tables, etc., as it goes, so in general you don't know what number something
% is going to have.  We'll refer to the "hairymath" section a little later.

\section{Trigonometric Parallax and Absolute Magnitude of RW Tri}

The modeling of the observations gives an HST relative parallax for
RW Tri of 2.59{$\pm$}0.29 mas.  The correction to absolute
parallax  adds
0.34 mas, giving an absolute parallax of 2.93{$\pm$}0.33 mas,
and a distance of   $341^{-31}_{+38}$ parsecs 
(Table~\ref{tbl-3}). This distance lies in the range of distances 
predicted from many other non-astrometric methods (Table~\ref{tbl-4}).

The distance modulus for RW Tri is 7.67.
Using Bruch and Engel's (1994) visual magnitude 
of 13.2  for the apparent magnitude
we obtain an absolute magnitude ($M_V$) of $5.53^{-0.22}_{+0.2}$. 
When using trigonmetric parallaxes to estimate the absolute
magnitude of a star, a correction should be made for the
Lutz-Kelker (LK) bias (\cite{Lutz73}).
Because of the galactic latitude and distance of RW Tri, 
and the scale height of the
stellar population of which it is a member,
we do not use a uniform density in space for calculating
the LK bias, but derive a density law that falls off as the 
-0.5 power of the distance at the distance of RW Tri.
This translates into $\it n $ = -3.5 as the power in the parallax distribution.
This $\it n$ is then used in an LK
algorithm modified by Hanson (H)(1979) to include the power law
of the
parent population. A correction of -0.12 $\pm$ 0.05 mag is derived for
the
LKH bias, which makes the absolute magnitude $5.41^{-0.23}_{+0.21}$.

Although RW Tri (Galactic coordinates: {\it l} = 147.03, {\it b} = -30.33) is 
well below the plane of the galaxy, 
our  proper motion determination gives it a velocity of only 15 km s$^{-1}$ relative
to the reference stars.  This is consistent with the 14 NLs with known
radial velocities all being less than 40 km s$^{-1}$ (\cite {Sha99}).  
This indicates ithat RW Tri is a member of 
the disk population of our galaxy.

\section{The Bailey Relation    }

In 1981, Bailey  presented a formula for surface 
brightness which has been used
to estimate the distances of CVs.
\begin{equation}
log(d) = \left(\frac{K}{5}\right) + 1 - \left(\frac{S_K}{5}\right) + 
log\left(\frac{R_2}{R_{\sun}}\right)
\end{equation}
where $\it d$ is the distance in parsecs, 
$\it K$ is the observed K magnitude, $S_K$ is the
surface intensity in the K-band, and $R_2$ is the radius of the
secondary. In 1994, Ramseyer examined  additional data and showed that
$S_K$ was not constant over a large range of $\it V-K$ as previously thought.
Using a $\it V-K$ of 3.08 (average of  $\it V$  at minimum of Smak (1995), 
Longmore et. al. (1981) and Walker (1963) minus the $\it K$  from Longmore et. al. 
(1981)),  and 
\begin{equation}
S_K = 4.35 + 0.022 * (\it{V-K})
\end{equation}
from Ramseyer (1994) Table 1 (Class V, $\it V-K$ 3.0-5.0), 
we get a ${S_K}$ of 4.42.  Using the mass-radius relation
from Warner (1995) 
\begin{equation}
R_2 = {M_2}^{(13/15)} 
\end{equation}
with Smak's (1995) secondary mass of 0.63,
we get  $R_2$ = 0.67.  Using these numbers in  Equation  3 we
derive a distance of 238 parsecs.  If the period-radius  relations of 
Warner (1995) or Patterson (1984) or the  mass-radius  relation of Smith and
Dhillon (1998)  are used  the distance estimate 
drops further to 217 parsecs. 

The infrared spectroscopic parallax,
when compared to the HST parallax,  shows
that  excess luminosity is present in the
K-band of these objects. This causes problems with the Bailey  method
of distance determination.    The distance may be underestimated in
Equation 3 due to the
assumption that the disc is not contributing to the K-band luminosity.

% The \notetoeditor{TEXT} command allows the author to communicate some
% information to the copy editor.  This information will appear as a 
% footnote on the printed copy for the aasms4 style file.  Nothing will 
% appear on the printed copy if the aaspp4 or aas2pp4 style file is used.

% In these sections, we see some additional math-related markup, and we
% have references to one of the tables (occurs later in the document)
% and the "hairymath" section immediately preceding this one.
%
% In the second paragraph, note the use of in-text math ($stuff$) including
% a couple of the miscellaneous symbol commands defined in the AASTeX macro
% package.
%
% This is the last section of the paper, so there is an \acknowledgments
% section at the end of the main body.

\section{Summary }

Trigonometric parallaxes can
provide distances which are independent of the assumptions (such as
the intrinsic absolute luminosity or shape of the spectral
energy distribution of the accretion disk)  that have been
used in the previous methods  of determining the distances for
CVs.
Our trigonometric parallax places RW Tri at a greater distance
than previously thought.  The more precise distance
reported here will improve knowledge of
the physical processes associated with this 
interesting object, placing it on an absolute scale.

%\placetable{tbl-1}
%\placetable{tbl-2}
%\placefigure{fig3}

\acknowledgments

This work is based on observation made with the NASA/ESA Hubble Space
Telescope, which is operated by  the Space Telescope Science Institute,
of  the Association of Universities for Research in Astronomy, Inc.,
under NASA contract NAS5-26555.
The HST Astrometry Science Team receives support through NASA grant
NAS5-1603.
Support for C.P.D. was provided by NASA through grant HF-1042.01-93A
awarded by the Space Telescope Science Institute, which is operated by
the Association of Universities for Research in Astronomy, Inc., for
NASA under contract NAS5-26555.
We thank Bill Spiesman, Cyndi Froning, Rob Robinson, Tom Harrison 
and Phil Ianna  for
helpful discussions and draft paper reviews.
  Denise Taylor and Lauretta Nagel provided 
assistance at the Space Telescope Science Institute.

\appendix

\clearpage

\begin{deluxetable}{crrrrrrrrrrr}
\footnotesize
\tablecaption{RW Tri and Reference Star Data \label{tbl-2}}
\tablewidth{0pt}
\tablehead{
\colhead{Star} & 
\colhead{$\xi$ (asec) \tablenotemark{a}}   &
 \colhead{$\eta$ (asec) \tablenotemark{a}} &
\colhead{$\sigma {_\xi}$ (mas)}   &
 \colhead{$\sigma {_\eta}$ (mas) }& 
 \colhead{V magnitude} &
 \colhead{Spectral Type}
}
\startdata
Ref 1   & 9.320 & -41.837 & 0.30    &   0.42 & 14.2 & G1 III\nl
Ref 2   & -271.562 & 42.123 &  0.54    &   0.72 & 14.1 & G0 II\nl
RW Tri &  30.600 & -88.424 &   0.45 &      0.72 & 13.2 & Novalike\nl
Ref 4   & 109.025 & -55.341 & 0.32 &      0.38 & 11.2 & A6 III\nl
Ref 5   &  -73.128 & 66.071 & 0.50 &      0.76 &13.8 & F5 Ib-II\nl
Ref 6   &  262.725 & -8.728 & 0.55 &      0.56 & 12.8 & G9 III\nl
\enddata
\tablenotetext{a}{$\xi$ and $\eta$ are relative positions }
\end{deluxetable}

\clearpage
 
\begin{deluxetable}{crrrrrrrrrrr}
\footnotesize
\tablecaption{Parallax and Proper Motion of RW Tri\label{tbl-3}}
\tablewidth{0pt}
\tablehead{
\colhead{Relative $\mu_\alpha$} &\colhead{Relative $\mu_\delta$} 
& \colhead{Relative Parallax}   &
 \colhead{Observed Ref. Parallax \tablenotemark{a}} &
    \colhead{Parallax}  &\colhead{Distance} \nl
\colhead{$(mas/yr^{-1})$}& \colhead{$(mas/yr^{-1})$} & \colhead{(mas)} & \colhead{(mas)} & \colhead{(mas)} 
& \colhead{(parsec)}  
}
\startdata
7.1 {$\pm$}0.4 &0.03 {$\pm$}0.7 &2.59{$\pm$}0.29  &0.34{$\pm$}0.16  &2.93{$\pm$}0.33 &$341^{-31}_{+38}$\nl
\enddata

% Text for table footnotes follows the tabular data and must be inside the
% deluxetable environment.  Note that it is OK to put \ref's in 
% \tablenotetext's.
 
\tablenotetext{a}{Correction to absolute parallax is derived from WIYN spectra}

\end{deluxetable}

\clearpage
 
\begin{deluxetable}{crrrrrrrrrrr}
\footnotesize
\tablecaption{Distance Estimates to RW Tri\label{tbl-4}}
\tablewidth{0pt}
\tablehead{
\colhead{Reference}    &
 \colhead{Distance in parsecs} & \colhead{Method} }
\startdata
Young \&Schneider (1981) & $>107$ &Non-detection of M1 dwarf - flux deficit in TiO Band\nl
Young \&Schneider (1981) & $>161$ &Non-detection of M4 dwarf - flux deficit in TiO Band\nl
Frank \& King (1981) &180 {$\pm$}70 &Best fit to the light curves in a quiescent state\nl
Borne (1977) &200 &Spectroscopic Parallax \nl
Young \&Schneider (1981) & $>224$ &Non-detection of M3 dwarf - flux deficit in TiO Band\nl
Warner   (1987) &224 &K-band magnitude \nl
Bailey   (1981) &247 &$\it V-K$ Surface Brightness Calibration \nl
Smak    (1995) &270 {$\pm$} 40 &Light Curve fitting\nl
Rutten et. al (1992)  &270 &Black body fit to spectrum of central part of disk \nl
Rutten et. al (1992)  &330  &Derived from fractional contribution of Secondary Star  \nl
HST  (1998) &$341^{-31}_{+38}$  &Trigonometric parallax, this paper\nl
Young \&Schneider (1981) & $>347$ &Non-detection of M1 dwarf - flux deficit in TiO Band\nl
Longmore et. al (1981) & 400 &Light curves at near IR wavelengths\nl
Horne \& Steining (1985) &500 &Modeling of Disk Properties from eclipse maps\nl
\enddata

\end{deluxetable}

\clearpage

\clearpage

\end{document}